\documentclass{aa}
\usepackage{graphicx}
\usepackage{natbib}
\usepackage{dcolumn}
\usepackage{subfigure}
\usepackage{txfonts}
\bibpunct{(}{)}{;}{a}{}{,}

\newcolumntype{d}[1]{D{.}{.}{#1}}

\begin{document}
\title{\object{AD Leonis}: Flares observed by XMM-Newton and Chandra} 
\author{E.J.M. van den Besselaar\inst{1,2} \and
  A.J.J. Raassen\inst{1,3} \and R. Mewe\inst{1} \and R.L.J. van der
  Meer\inst{1} \and M. G\"udel\inst{4} \and M. Audard\inst{5} } 

\offprints{E.J.M. van den Besselaar, \\
  \email{besselaar@astro.kun.nl}} 

\institute{
  SRON National Institute for Space Research, Sorbonnelaan 2, 3584 CA
  Utrecht, The Netherlands \\  
  \email{A.J.J.Raassen@sron.nl; R.Mewe@sron.nl;
  R.L.J.van.der.Meer@sron.nl} 
\and
  Department of Astrophysics, University of Nijmegen, P.O. Box 9010,
  6500 GL Nijmegen, The Netherlands  
\and 
  Astronomical Institute "Anton Pannekoek", Kruislaan 403, 1098 SJ
  Amsterdam, The Netherlands
\and
  Paul Scherrer Institut, W\"urenlingen \& Villigen, 5232 Villigen
  PSI, Switzerland \\ \email{guedel@astro.phys.ethz.ch}
\and
  Columbia Astrophysics Laboratory, Columbia University, 550 West
  120th Street, New York, NY 10027, USA \\
  \email{audard@astro.columbia.edu} 
}
\date{Received 3 February 2003 / Accepted 2 September 2003}

\abstract{The M-dwarf \object{AD Leonis} has been observed with the
  Reflection Grating Spectrometers and the European Photon Imaging
  Camera aboard XMM-Newton and also with the Low Energy Transmission
  Grating Spectrometer aboard the Chandra X-ray Observatory. In the
  observation taken with XMM-Newton five large flares produced by
  \object{AD Leo} were identified and only one in the observation
  taken with Chandra. A quiescent level to the lightcurves is
  difficult to define, since several smaller flares mutually overlap
  each other. However, we defined a quasi-steady state outside of
  obvious flares or flare decays. The spectra from the flare state and
  the quasi-steady state are analysed separately. From these spectra
  the temperature structure was derived with a multi-temperature model
  and with a differential emission measure model. The
  multi-temperature model was also used to determine the relative
  abundances of \element{C}, \element{N}, \element{O}, \element{Ne},
  \element{Mg}, \element{Si}, \element{S}, and
  \element{Fe}. \element{He}-like ions, such as \ion{O}{vii} and
  \ion{Ne}{ix}, produce line triplets which are used to determine or
  constrain the electron temperature and electron density of the
  corresponding ion. During the flare state a higher emission measure
  at the hottest temperature is found for both XMM-Newton and Chandra
  observations. The derived abundances suggest the presence of an
  \textit{inverse} First Ionization Potential effect in the corona of
  \object{AD Leo}. 
}

\maketitle

\section{Introduction}

\object{AD Leo} is an M-dwarf with spectral type M3.5~V at a distance
of 4.7~parsec\footnote{SIMBAD Database:
  http://simbad.u-strasbg.fr/sim-fid.pl}. Many cool stars (F -- M)
maintain active coronae with temperatures up to 20~MK. Our goal is to
determine differences in the physical coronal conditions such as
temperatures, emission measures, abundances, and densities between
different states of the corona of \object{AD Leo}. We note that in the
corona of the \object{Sun} a First Ionization Potential (FIP) effect
is observed \citep{fip} which implies that elements with a low FIP
(say $\la$~10~eV) are enhanced in the corona relative to the
photospheric values. But for other active stars an \textit{Inverse}
FIP (IFIP) effect was suggested \citep{ifip1, ifip2}. The underlying
mechanism for these FIP and IFIP effects is not well understood. In
this paper, the abundances are measured to see if there are anomalies
and whether these are different for the flare state and quasi-steady
state. 

\object{AD Leo} is known to be capable of frequent flaring in the
X/EUV/optical regime \citep{kahn, flare, exosat, uvopt, nh}. Flares on
M-dwarfs may play a part in the heating mechanism of the outer
atmospheres of stars. \citet{latetypeflare}, \citet{flareheating}, and
\citet{flareheating2} describe this mechanism. They suggest that the
'quiescent' emission is in fact a superposition of multiple small
flares. The expression 'quasi-steady' is therefore used in this paper
to refer to the state between the distinct flares.  

\begin{figure*}
\centering
  \includegraphics[angle=270, totalheight=5cm,width=17cm]{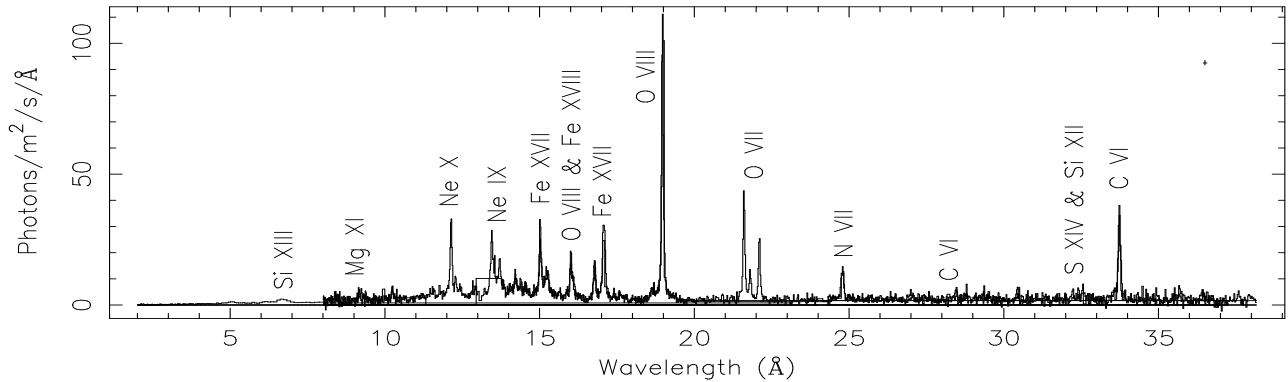}   
 \caption{The X-ray spectra taken by \textsc{rgs 1}, \textsc{rgs 2}
          and \textsc{epic-mos 2} aboard XMM-Newton. The dominant
          lines in the spectrum have been labeled with the
          corresponding ion. \textsc{epic-mos 2} has a lower
          resolution and is used only from $\sim$2 to
          $\sim$14~\AA. \textsc{rgs} 1 and 2 have a higher resolution
          and are used from 8 to 38~\AA. A typical error bar for the
          continuum is included in the upper right corner.}
  \label{fig:xmm:spectrum}
\end{figure*}

\begin{figure*}
\centering
  \includegraphics[angle=270, totalheight=5cm,width=17cm]{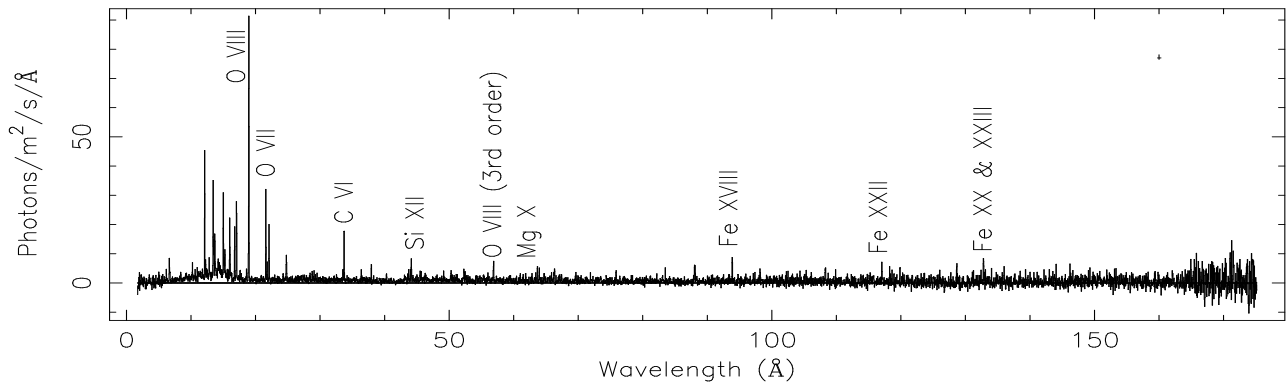} 
  \caption{The X-ray spectrum taken by \textsc{letgs} aboard
           Chandra. The dominant lines in the spectrum have been
           labeled with the corresponding ion. The labels for lines
           below 20~\AA~can be found in Fig.~\ref{fig:xmm:spectrum}. A
           typical error bar for the continuum is included in the
           upper right corner.} 
  \label{fig:chandra:spectrum}
\end{figure*}

Here we investigate the high-resolution spectra of \object{AD Leo}
taken by the XMM-Newton satellite and the Chandra X-ray
Observatory. The outline of this paper is the following. First, both
observations are discussed. Section~\ref{sec:datared} gives the
procedures followed to reduce the data. Section~\ref{sec:xmm:flare}
describes the data analysis of the XMM-Newton observation during the
flaring and quasi-steady states, whereas Sect.~\ref{sec:chandra:flare}
describes the analysis for the Chandra observation. In
Sect.~\ref{sec:discussion} we discuss our results, and finally in
Sect.~\ref{sec:conclusion} our conclusions are given. 

\section{Observations}

On May 14, 2001 \object{AD Leo} was observed by XMM-Newton. Here we
investigate the spectra taken with the two Reflection Grating
Spectrometers, \textsc{rgs}, and the European Photon Imaging Camera,
\textsc{epic-mos 2}. (\textsc{epic-mos 1} was in timing mode,
\textsc{epic-pn} has less spectral resolution than the
\textsc{epic-mos}.) The spectral resolution of \textsc{epic-mos 2} is
lower than that of \textsc{rgs}, but its sensitivity is higher. The
exposure time was about 36~ks. The state of the instruments during
this observation is given in Table~\ref{tab:xmm:state} together with
the total observation time of the corresponding instrument. In this
observation no strong solar flares were detected and therefore no time
intervals were rejected.   

\begin{table}
\centering
\caption{The state of the instruments aboard XMM-Newton together with
         the corresponding observation times in seconds.} 
\label{tab:xmm:state}
 \begin{tabular}{l l l l}
  \hline
\hline
  Instr. & Mode & Filter & Obs. Time \\
  \hline
  \textsc{epic-mos 2} & large window & thick 		& 35\,824 \\
  \textsc{rgs 1} & spectro + q & not applicable 	& 36\,354 \\
  \textsc{rgs 2} & spectro + q & not applicable 	& 36\,354 \\
  \hline
 \end{tabular}
\end{table}

The spectrum of this dataset is shown in
Fig.~\ref{fig:xmm:spectrum}. The wavelength range used for
\textsc{rgs} is 8 to 38~\AA~and the spectral line resolution is
$\Delta\lambda\sim$~0.07~\AA~at full width at half maximum
(FWHM). \textsc{rgs} has a wavelength accuracy of $\sim$8~m\AA~and a
maximum effective area of about 140~cm$^{2}$ around
15~\AA~\citep{rgs}. The wavelength range used for \textsc{epic-mos 2}
was constrained to $\sim$1.8 to $\sim$14~\AA~because of the better
resolution of \textsc{rgs} at longer wavelengths. The background
(about 1~photon m$^{-2}$ s$^{-1}$ \AA$^{-1}$) has been subtracted from
these spectra. \object{AD Leo} was also ob\-ser\-ved, half a year
ear\-lier, with the Low Ener\-gy Trans\-mis\-sion Gra\-ting
Spec\-tro\-me\-ter (\textsc{letgs}) aboard Chandra. This instrument
has taken the X-ray spectrum on October 24, 2000 during an exposure
time of 48.41~ks. The background-subtracted spectrum is shown in
Fig.~\ref{fig:chandra:spectrum}. The background consists of about
1~photon m$^{-2}$ s$^{-1}$ \AA$^{-1}$ below 60~\AA~and steadily
increasing to about 8~photons~m$^{-2}$~s$^{-1}$~\AA$^{-1}$ at
$\sim$170~\AA.   

The \textsc{letgs} consists of an imaging camera (\textsc{hrc-s}) with
a grating spectrometer (\textsc{letg}). At FWHM, this instrument has a
spectral resolution of $\Delta\lambda\sim0.06$~\AA~\citep{chandra} and
a wavelength uncertainty of a few m\AA~below 30~\AA, increasing to
about 0.02~\AA~above. The wavelength range is 1 to 175~\AA.  

The number of photons m$^{-2}$ s$^{-1}$ \AA$^{-1}$ for the lines in
the XMM-Newton spectra show a small difference from those of the
Chandra spectra. We believe that the difference is due to different   
activity levels between the XMM-Newton and Chandra observations.

\section{Data reduction}
\label{sec:datared}

The spectra taken with the XMM-Newton satellite are pipelined using
the XMM-Newton SAS, ''Science Analysis System'', version 5.3.3. This
program reads the current calibration files of the instruments and the
observation data files. The relevant information in these files is
written to new files which contain the source spectrum, the background
spectrum, and the response matrix with information about e.g., the
line-spread-function and the effective area of each instrument
used. To construct the \textsc{rgs} spectrum 95~\% of the
cross-dispersion function of the data is used. In our analysis we have
used only the first order spectrum. 

To extract the spectrum of the \textsc{epic-mos 2} data a circle with
a radius of 40\arcsec~around the image of \object{AD Leo} is
taken. The response matrix and ancillary files, containing the
effective area, are generated with SAS. To determine the background
level, a circle with identical radius is used in a part of the
detector without any sources. 

The Chandra data is pipelined using the CIAO 2.2 program (Chandra
Interactive Analysis of Observations). The spectrum, response matrix,
and effective area files are determined by a local extracting
software, using a bow-tie shaped region around the dispersion axis as
extraction region for the spectrum. In our analysis of this
observation we have co-added the +1 order and -1 order spectra. The
background level is determined from a rectangular box 10\arcsec~--
40\arcsec~above and below the dispersion axis on the detector, to get
the best statistics on the background. The background spectrum is then
scaled per bin from the rectangular box to the size of the bow-tie in
the corresponding wavelength bin.  

We use the software package SPEX \citep{spex} in combination with the 
updated MEKAL code \citep{mekal} for an optically thin plasma in 
Collisional Ionization Equilibrium 
(CIE)\footnote{http://www.sron.nl/divisions/hea/spex/version1.10/line/} 
to analyse the X-ray spectra of XMM-Newton and Chandra.

\section{Data analysis of the XMM-Newton observation}
\label{sec:xmm:flare}

\begin{figure}
\centering
\includegraphics[angle=270,width=\linewidth]{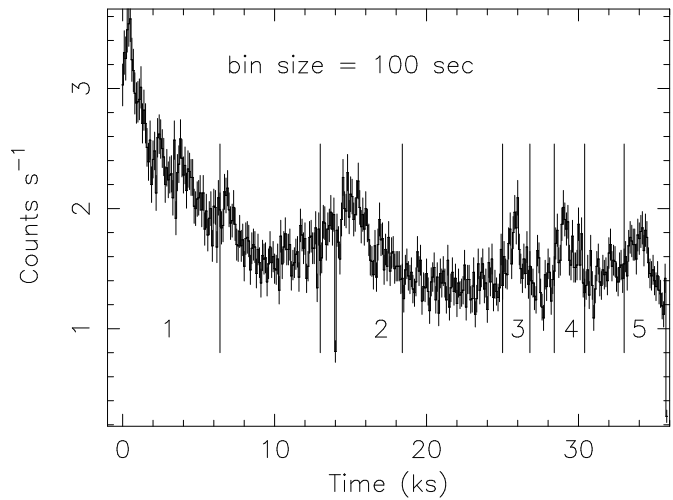}
  \caption{Lightcurve of the observation taken by \textsc{epic-mos
  2}.} 
  \label{fig:xmm:lightcurve}
\end{figure}

In the lightcurve of the X-ray observation of \object{AD Leo} taken
with \textsc{epic-mos 2} five flares are distinguished. The time
intervals of the flares are given in Table~\ref{tab:xmm:flares} and
shown in Fig.~\ref{fig:xmm:lightcurve}. We have analysed the summed
time intervals of the flares (17.2 ks). Indeed, the signal-to-noise
ratio of individual flares is too low to obtain clear results. It is
difficult to clearly distinguish quiescent emission (if that exists)
from the overlapping flare events. It is possible that no physical
quiescent emission is present at all in the corona of \object{AD Leo}, 
but that the emission is produced by a superposition of multiple small 
flares \citep{latetypeflare,flareheating,flareheating2}.

\begin{table}
\centering
 \caption{Time intervals of the flares in seconds after the start of
 the XMM-Newton observation (t$_0=52043.8667$ MJD).}  
 \label{tab:xmm:flares}
  \begin{tabular}{l l  l}
   \hline
   \hline
   Nr. & Begin & End \\
   \hline
   1 & 0 	& 6\,400 \\
   2 & 13\,400 & 18\,400 \\
   3 & 25\,400 & 27\,400 \\
   4 & 28\,900 & 30\,700 \\
   5 & 33\,400 & 35\,400 \\ 
   \hline
   \multicolumn{2}{l}{Total flare state:}  & 17.2 ks \\
   \multicolumn{2}{l}{Total quasi-steady state:} & 18.8 ks \\
   \hline
  \end{tabular}
\end{table}

The temperature structure in the corona is determined in two different
ways. First, we have measured the temperatures with a
multi-temperature model using the MEKAL code. This model is
simultaneously fitted to the \textsc{rgs 1}, \textsc{rgs 2}, and
\textsc{epic-mos 2} spectra. The temperatures, emission measures, and
abundances of several elements were derived from this model. The
second method to determine the temperature structure is by use of a
differential emission measure (DEM) model based on the polynomial
method \citep{dem}. The used interstellar hydrogen column density  
(N$_{\rm H}$) is $3\cdot10^{18}$ cm$^{-2}$ \citep{nh}.

\begin{table*}
\centering
 \caption{Temperatures, emission measures, and coronal abundances of
           several elements as determined from the spectrum during the
           flares and quasi-steady state. The abundance ratios are
           relative to solar photospheric abundances (Col. 6;
           $\log{\rm{A}_{\rm{H}}}=12$). The numbers in parentheses are
           the statistical 1$\sigma$ uncertainties. The X-ray
           luminosity (\textit{L}$_\textrm{x}$) is measured between
           0.3 and 10 keV. \textit{EM}$_{\rm total}$ is the total of
           the three given emission measures. Cols. 2 and 3 give the
           results of the XMM-Newton observation, discussed in
           Sect.~\ref{sec:xmm:temp}, and Cols. 4 and 5 those of the 
           Chandra observation which are discussed in
           Sect.~\ref{sec:chandra:temp}.}   
  \label{tab:temp}
\begin{tabular}{l r@{.}l l r@{.}l l |  r@{.}l l r@{.}l l | l l}
\hline
\hline
& \multicolumn{6}{c|}{XMM-Newton} & \multicolumn{6}{c|}{Chandra} & 
\multicolumn{2}{c}{Solar Photospheric}\\
Parameter 				& \multicolumn{3}{c}{Flare}
& \multicolumn{3}{c|}{Quasi-steady} & \multicolumn{3}{c}{Flare}
& \multicolumn{3}{c|}{Quasi-steady} & \multicolumn{2}{c}{Abundance} \\
\hline
\textit{T}$_1$ (MK) 		& 2&88 & (0.06) & 2&84 & (0.06) & 
2&7 & (0.1) & 2&8 & (0.1) & & \\
\textit{EM}$_1$ (10$^{51}$ cm$^{-3}$)& 1&2 & (0.1)& 1&2 & (0.1) &
0&8 & (0.1) & 0&54 & (0.07) & & \\
\textit{T}$_2$ (MK) 		& 7&1 & (0.1) 	& 7&1 & (0.1) &
7&1 & (0.3) & 6&7 & (0.2) & & \\
\textit{EM}$_2$ (10$^{51}$ cm$^{-3}$)& 2&0 & (0.2) & 1&5 & (0.1) &
1&2 & (0.2) & 1&0 & (0.1) & & \\
\textit{T}$_3$ (MK) 		& 20&2 & (1.0) 	& 20&1 &  (1.5) &
17&6 & (1.5) & 14&8 & (1.9) & & \\
\textit{EM}$_3$ (10$^{51}$ cm$^{-3}$)& 0&92 & (0.07) & 0&46& (0.05) &
1&3 & (0.1) & 0&37 & (0.06) & & \\
\textit{L}$_\textrm{x}$ (10$^{28}$ erg/s) & 4&6 &	& 3&5 &    & 
 4&4 & & 2&3 & & & \\
Abun. \element{C}/\element{O} 		& 1&6 & (0.2) 	& 1&7 & (0.2) &
 1&7 & (0.5) & 1&8 & (0.3) & $\log{\rm{A}_{\rm{C}}}$ &8.39\\
Abun. \element{N}/\element{O} 		& 1&0 & (0.1) 	& 0&9 & (0.1) &
 1&1 & (0.3) & 0&8 & (0.2) &  $\log{\rm{A}_{\rm{N}}}$ & 8.05\\
Abun. \element{O}/\element{O} 		& 1&00 & (0.09) & 1&0 & (0.1) &
 1&0 & (0.2) & 1&0 & (0.1) &  $\log{\rm{A}_{\rm{O}}}$ &8.69\\
Abun. \element{Ne}/\element{O} 		& 1&1 & (0.1) 	& 1&2 & (0.1) &
 1&4 & (0.3) & 1&2 & (0.2) & $\log{\rm{A}_{\rm{Ne}}}$ & 8.09\\
Abun. \element{Mg}/\element{O}		& 0&48 & (0.07)	& 0&54 &(0.09) &
 0&5 & (0.2) & 0&21 & (0.08) &  $\log{\rm{A}_{\rm{Mg}}}$ & 7.58\\
Abun. \element{Si}/\element{O}		& 0&8 & (0.1)	& 0&9 & (0.1) &
 0&8 & (0.2) & 1&0 & (0.1) &  $\log{\rm{A}_{\rm{Si}}}$ & 7.55\\
Abun. \element{S}/\element{O}		& 0&6 & (0.1)	& 0&6 & (0.1) &
 0&3 & (0.1) & 0&3 & (0.1) &  $\log{\rm{A}_{\rm{S}}}$ & 7.21\\
Abun. \element{Fe}/\element{O} 		& 0&46 & (0.05)	& 0&48 &(0.05) &
 0&44 & (0.09) & 0&35 & (0.04) &  $\log{\rm{A}_{\rm{Fe}}}$ & 7.50\\
Abun. \element{O}/\element{H}		& 0&74 & (0.05)	& 0&71 & (0.05) &
 1&0 & (0.1) & 1&1 & (0.1) & \\
\element{O}/\element{H}$\cdot$\textit{EM}$_{\rm total}$ & 3&1 & (0.3) & 2&2 & (0.2) 
                & 3&3 & (0.4) & 2&1 & (0.2) & & \\
  \hline
  $\chi^2$/d.o.f. & \multicolumn{3}{c}{5598/4574} & \multicolumn{3}{c|}{5529/4574} &
  \multicolumn{3}{c}{1700/1428} & \multicolumn{3}{c|}{5271/4532} & \\
  \hline
 \end{tabular}
\end{table*}

\subsection{Temperatures and abundances}
\label{sec:xmm:temp}

A multi-temperature fit to the data was performed. From this model
three temperatures and their corresponding emission measures were
determined, together with the abundances of \element{C}, \element{N},
\element{O}, \element{Ne}, \element{Mg}, \element{Si}, \element{S},
and \element{Fe}. The $\chi^2$ statistics did not improve when
additional components were added. The results from this analysis are
given in Table~\ref{tab:temp}.  

The uncertainties in this table are 1$\sigma$ statistical
uncertainties only. However we caution that additional systematic
uncertainties exist. For example, a typical 10\% uncertainty is
present in the atomic database used to model the lines. Therefore
abundances and EMs are affected by this uncertainty as well. Typically
their total uncertainties will range from 15 to 20\%. 

Unfortunately, the continuum is not well constrained and the absolute
abundances (relative to \element{H}) are thus not uniquely
derived. Indeed, if the continuum level is difficult to measure, the
EMs cannot be easily constrained. Incidentally any line flux can be
modeled with any product of the elemental abundance and the EM, e.g.,
\element{O}/\element{H}$\cdot$\textit{EM}$_{\rm total}~\cong$
constant. On the other hand, an abundance ratio (e.g., relative to 
\element{O}) is rather stable and allow us to compare between the
XMM-Newton and Chandra observations.

The abundances are relative to their corresponding solar photospheric
values from \citet{anders}, except for \element{O}, \element{C}, and
\element{Fe} for which more recent values were used
\citep{oxygen,carbon,iron}.  

To study differences between the flare state and quasi-steady state,
the latter has been analysed as well. The results from this analysis
are also given in Table~\ref{tab:temp}. The emission measure of the
two highest temperatures is significantly lower for the quasi-steady
state than for the flare state. The abundances, however, do not show a
significant difference between the flares and the quasi-steady state.

\subsection{DEM modeling}
\label{sec:xmm:dem}

\begin{figure}
\centering
   \includegraphics[angle=270, width=\linewidth]{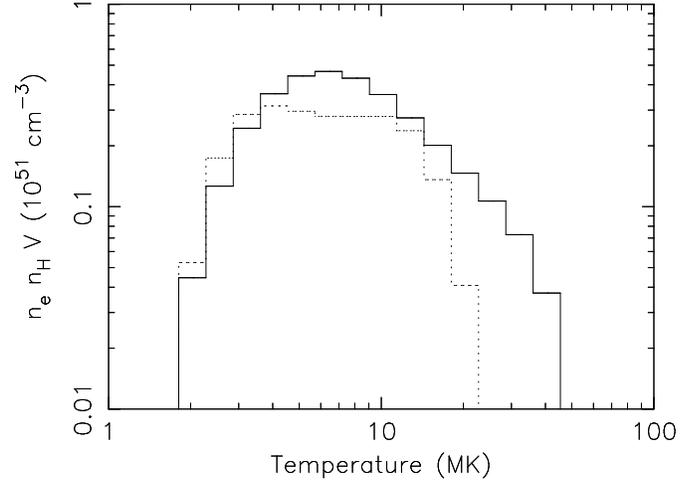}
 \caption{The differential emission measure, as obtained from the
          polynomial method with order eight, during the flare state
          (solid line) and quasi-steady state (dotted line) of
          \object{AD Leo} as measured with XMM-Newton.} 
 \label{fig:xmm:dem} 
\end{figure}

To determine a smoother temperature distribution a DEM modeling is
performed for the two data sets consisting of the flare state and
quasi-steady state. \citet{dem} describe different methods to
determine the differential emission measure distribution which are
based on a clean, polynomial, or regularization method. The polynomial
method with order eight is used here to derive the emission measure
distribution of these spectra, but the results of the other methods
are in agreement. The abundances derived from fitting them
simultaneously with the emission measure distribution agree very well
with those received from the multi-temperature fit. The derived
distribution is shown in Fig.~\ref{fig:xmm:dem} for both
states. During the flare state the emission measure of the highest
temperature range is higher.  

\subsection{Analysis of He-like ions}
\label{sec:xmm:triplet}

The Helium-like ions, such as \ion{O}{vii} and \ion{Ne}{ix}, produce
the well-known line triplets in the spectrum. Such a triplet consists
of a resonance ($r$), forbidden ($f$), and an intercombination ($i$)
line. The flux ratios of these lines are dependent on the electron
temperatures and densities. A higher density results in weaker
forbidden lines and stronger intercombination lines. Therefore the
line flux ratio $f/i$ is a density sensitive quantity.

The \ion{O}{vii} line triplet is located between 21.6 and
22.1~\AA. The fluxes of these lines have been measured during both
states of \object{AD Leo}. The values are given in
Table~\ref{tab:tripletflux}. In the multi-temperature model the
abundance of \element{O} is set to zero and the other abundances,
temperatures, and emission measures are fixed at the values given in
Table~\ref{tab:temp}. The effect to the continuum by setting the
abundance of \element{O} to zero is negligible. Three delta functions
are added on top of this model to measure the contribution of
\element{O} to these lines in a small wavelength range around the
\element{O} lines. These delta functions are convolved with the
instrumental line profile and their amplitudes are fitted to the
observations. With this procedure we have corrected for possible
blends by other lines at these wavelengths and only the contribution
of \element{O} is measured. For the \ion{Ne}{ix} line triplet the same
procedure is applied. The effect to the continuum by setting
\element{Ne} to zero is negligible as well. Due to detector failure
the \ion{O}{vii} lines could only be measured in the spectrum taken
with \textsc{rgs 1} and the \ion{Ne}{ix} lines in the spectrum of
\textsc{rgs 2} only. The fluxes for the \ion{Ne}{ix} lines are also
given in Table~\ref{tab:tripletflux}. The densities derived from
\ion{Ne}{ix} are less reliable due to blending by several \element{Fe}
lines. 

\begin{table*}
\centering
\caption{Fluxes in ph s$^{-1}$ m$^{-2}$ measured at Earth for some
         \element{He}-like lines as determined from \textsc{rgs} 1
         (\ion{O}{vii}) and 2 (\ion{Ne}{ix}) are given in Cols. 3 and
         4, while the fluxes from the \textsc{letgs} are given in
         Cols. 5 and 6. (See Sects.~\ref{sec:xmm:triplet} and 
         \ref{sec:chandra:triplet} for more information.)}
 \label{tab:tripletflux}
\begin{tabular}{l l l l| l l}
  \hline
  \hline
  & & \multicolumn{2}{c|}{XMM-Newton} & \multicolumn{2}{c}{Chandra} \\
  Ion 			& $\lambda$(\AA) & Flare 	& Quasi-steady 
    & Flare & Quasi-steady\\
  \hline
  \ion{O}{vii} ($r$) & 21.59	& 3.7 (0.4) & 3.7 (0.4)	&3.6 (0.7)& 3.3 (0.4) \\
  \ion{O}{vii} ($i$) & 21.80	& 1.0 (0.2) & 0.8 (0.3) &1.3 (0.5)& 0.7 (0.2) \\
  \ion{O}{vii} ($f$) & 22.09	& 1.7 (0.3) & 2.2 (0.2) &3.1 (0.6)& 2.0 (0.3) \\
  \hline
  \ion{Ne}{ix} ($r$) & 13.44	& 2.6 (0.3) & 1.7 (0.3) &2.1 (0.4)& 1.9 (0.2) \\
  \ion{Ne}{ix} ($i$) & 13.53	& 0.4 (0.3) & 0.4 (0.3) &0.4 (0.3)& 0.5 (0.2) \\
  \ion{Ne}{ix} ($f$) & 13.69 	& 1.5 (0.3) & 1.1 (0.2) &1.5 (0.4)& 1.2 (0.2) \\
  \hline
 \end{tabular}
\end{table*}

With the use of two flux ratios ($G=(f+i)/r$ and $R=f/i$) in
combination with the tables of \citet{porquet}, the electron
temperatures and densities are derived for the corresponding ions. The
results are presented in Table~\ref{tab:tempdens}. The electron
temperature and upper limits of the electron density for \ion{O}{vii}
are higher during the flare state as compared to the quasi-steady
state, but the difference is not formally significant. These
temperatures and densities are representative for those plasmas in
which the \element{He}-like lines are formed, and form only part of
the range presented in Fig.~\ref{fig:xmm:dem}. 

\begin{table*}
\centering
   \caption{The electron temperatures $T_{\rm e}$ (MK) and electron
            densities $n_{\rm e}$ (cm $^{-3}$) as derived from the
            tables of \citet{porquet} for the XMM-Newton observation
            (Cols. 2 and 3, Sect.~\ref{sec:xmm:triplet}) and for the
            Chandra observation (Cols. 4 and 5,
            Sect.~\ref{sec:chandra:triplet}). The numbers in
            parentheses are the 1$\sigma$ uncertainties, and the upper
            and lower limits are 1$\sigma$ limits.}  
   \label{tab:tempdens}
 \begin{tabular}{ l l l| l l}
  \hline
  \hline
& \multicolumn{2}{c|}{XMM-Newton} & \multicolumn{2}{c}{Chandra}\\
  Ion 		& $T_{\rm e}$ (flare) & $T_{\rm e}$ (quasi-steady) 
        & $T_{\rm e}$ (flare) & $T_{\rm e}$ (quasi-steady)   \\
  \hline
  \ion{O}{vii} & 3 (1) 	        & 2.5 (1.0) &$<2.5$ 	     & 2.5 (1)\\
  \ion{Ne}{ix} & 4 (2)		& 3.0 (1.2) & $<6.5$         & $<5$  \\
  \hline
  \\
  Ion 		& ${n}_{\rm e}$ (flare) & ${n}_{\rm e}$ (quasi-steady) 
          & ${n}_{\rm e}$ (flare) & ${n}_{\rm e}$ (quasi-steady)  \\ 
  \hline
  \ion{O}{vii} 	& 3$\cdot10^{10}<{n}_{\rm e}<1\cdot10^{11}$ 
              & $<3\cdot10^{10}$ & $1\cdot10^9<n_{\rm e}<7\cdot10^{10}$ & $<3\cdot10^{10}$\\
  \ion{Ne}{ix} 	& $<3\cdot10^{12}$ & $< 3.5\cdot10^{12}$ & $<3.5\cdot10^{12}$	& $<7\cdot10^{11}$\\
  \hline
 \end{tabular}
\end{table*}

\section{Data analysis of the Chandra observation}
\label{sec:chandra:flare}

\begin{figure}
\centering
\includegraphics[angle=270,width=\linewidth]{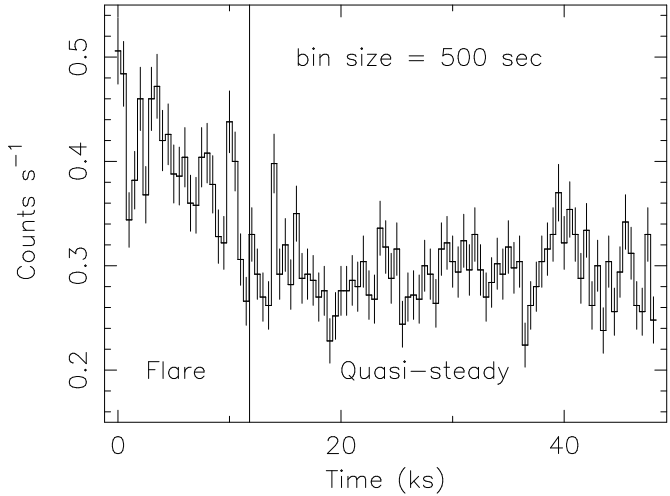} 
  \caption{The zeroth order lightcurve of the observation taken by
    \textsc{letgs}.} 
  \label{fig:chandra:lightcurve}
\end{figure}

The lightcurve from the \textsc{letgs} spectrum is produced by taking
the counts in a small circle around the zeroth order image of
\object{AD Leo} with a radius of 1.4\arcsec. From this lightcurve,
Fig.~\ref{fig:chandra:lightcurve}, we conclude that one flare was
produced by \object{AD Leo} at the beginning of the observation
(t$_0=51841.6286$ MJD). This flare lasted for 11.8~ks. The remainder
of the dataset (37.5~ks) is used to describe the quasi-steady
state. As mentioned before, we cannot clearly speak of quiescent
emission. 

Just like for the XMM-Newton spectra, a multi-temperature fit and a
DEM model were performed for the data to derive the temperature
structure.

\subsection{Temperatures and abundances}
\label{sec:chandra:temp}
 
A multi-temperature fit is performed to the data based on CIE
plasmas. From this model three temperatures were derived together with
the corresponding emission measures. These results and the derived
abundances are given in Table~\ref{tab:temp} together with the values
for the XMM-Newton observation. For the same reason as in
Sect.~\ref{sec:xmm:temp} the abundances are normalized to
\element{O}.

\subsection{DEM modeling}

\begin{figure}
\centering
 \includegraphics[angle=270,width=\linewidth]{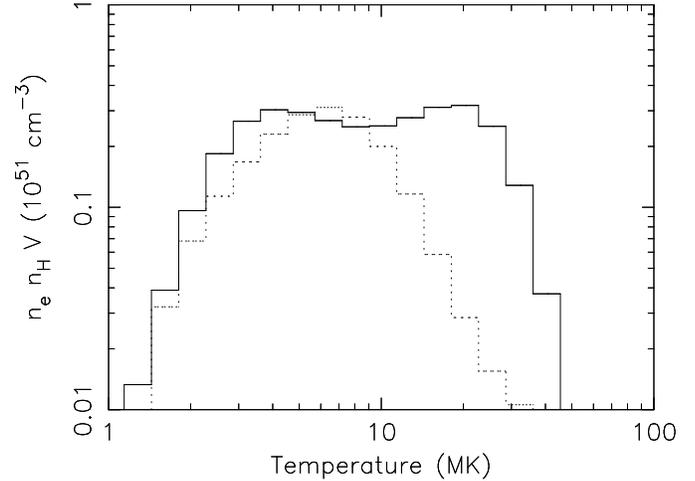}
 \caption{The differential emission measure, as obtained from the
          polynomial method with order eight, during the flare state
          (solid line) and quasi-steady state (dotted line) of
          \object{AD Leo} as measured with Chandra.}
 \label{fig:chandra:dem} 
\end{figure}

From the flare spectrum and quasi-steady spectrum a differential
emission measure distribution is derived as shown in
Fig.~\ref{fig:chandra:dem}. This emission measure distribution is
produced by the polynomial method with order eight. From this figure
we conclude that the high temperature range has a larger contribution
to the flare spectrum compared with the quasi-steady state. The other
temperatures have roughly the same contribution. Also in this case,
the results from the other methods agree very well with the DEM
distribution and abundances already given in this paper.

\subsection{Analysis of He-like line triplets}
\label{sec:chandra:triplet}

In the spectrum taken with Chandra, the line triplets produced by
\element{He}-like ions are present as well. We have measured the
fluxes of the \ion{O}{vii} and \ion{Ne}{ix} lines in the same way as
the triplets in the XMM-Newton spectra. The fluxes for both the flare
and quasi-steady state are given in Table~\ref{tab:tripletflux}
together with the fluxes as measured from the XMM-Newton
observation. In this spectrum also the line triplet produced by
\ion{Si}{xiii} between 6.6 and 6.8~\AA~is present, but the fluxes of
these lines are not significant enough to obtain a constraint on the
electron temperature and density. 

From the fluxes of these \element{He}-like triplets the electron
temperatures and electron densities of the respective ions were
derived using the tables of \citet{porquet}. The obtained temperatures
and densities are given in Table~\ref{tab:tempdens}. These
temperatures and densities are representative for those plasmas in
which the \element{He}-like lines are formed, and form only part of
the range presented in Fig.~\ref{fig:chandra:dem}.

\section{Discussion}
\label{sec:discussion}

\subsection{Flaring versus quasi-steady state}

The flare state and quasi-steady corona of the M-dwarf \object{AD Leo}
have been analysed in both the XMM-Newton and Chandra
spectra. Therefore the results are compared to determine the
conditions during a flare and the state between the flares. As
mentioned before, the quasi-steady emission is not clearly the
quiescent emission, if such exists, but can possibly be identified
with a superposition of decaying flares, and additional smaller
flares.  

From the DEM model, a difference between the flare and quasi-steady
state can be seen for both observations. The relatively low
($\sim$3~MK) and intermediate ($\sim$7~MK) temperature ranges are
present in almost the same amounts in both states. The difference
appears in the emission measure of the highest temperature range (20
-- 30~MK). The contribution of these temperatures is higher during the
flare state. This is just as expected for a large flare. Smaller
flares get less hot, and therefore the emission measure of the high
temperature range is smaller during the quasi-steady state. Also the
electron temperatures derived from the \element{He}-like line triplets
are slightly higher during a flare.

\subsection{Abundances}
\label{sec:abun}

\begin{figure}
\centering
   \includegraphics[angle=270, width=\linewidth]{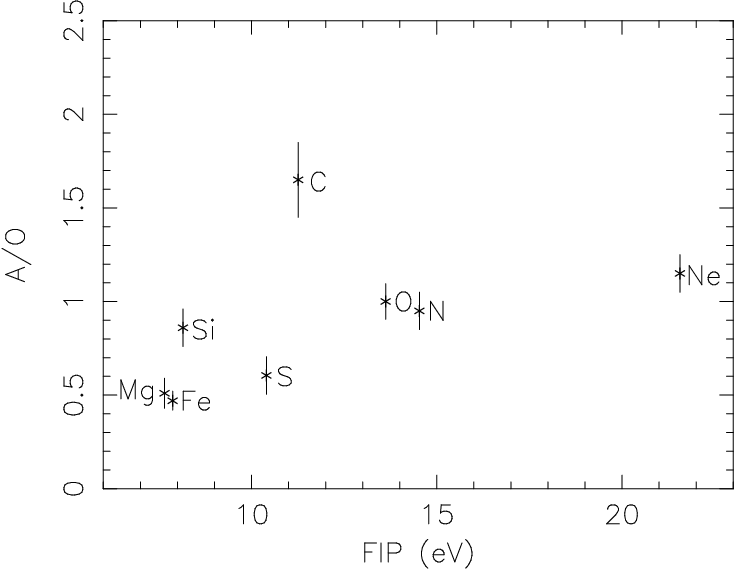}
   \includegraphics[angle=270, width=\linewidth]{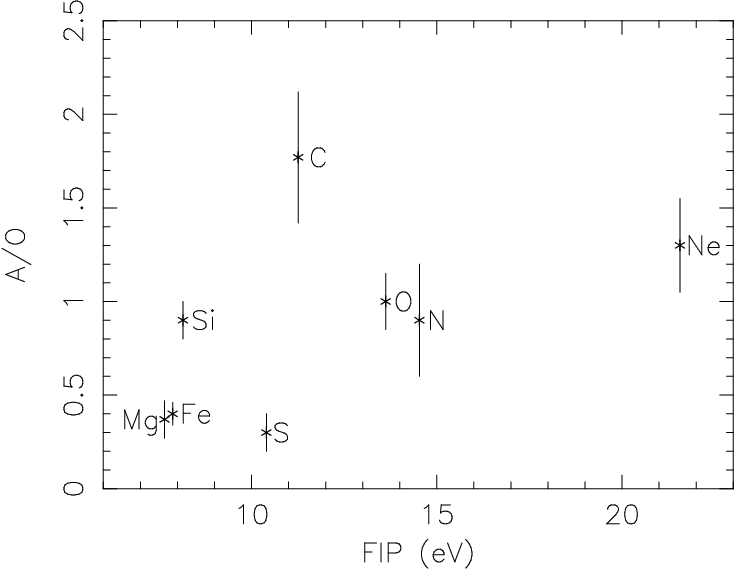}
 \caption{Abundances versus the corresponding first ionization
   potentials as measured with XMM-Newton (top) and Chandra (bottom).}
 \label{fig:fip} 
\end{figure}

The corona of the \object{Sun}, which is less active than the corona
of \object{AD Leo}, shows a first ionization potential effect. As
explained in Sect.~\ref{sec:xmm:temp} the absolute abundance
measurements are very uncertain, and therefore relative abundances are
given. There is no significant difference in the relative abundances
of the different elements during the flare state and the quasi-steady
state. Therefore the average coronal abundances are plotted in
Fig.~\ref{fig:fip} against the first ionization potential (FIP) of the
corresponding element for the XMM-Newton observations as well as for
the Chandra observation. Elements with a low FIP ($<$~10 eV) are
somewhat depleted, instead of enhanced, with respect to the solar
photospheric values. 

Despite the difficulty to obtain reliable absolute abundances (because
the continuum is poorly determined and thus the absolute level of the
EM may be equally uncertain), relative abundances show a clear pattern
which will allow us to compare that to the pattern of the \object{Sun}
and other (active) stars. Our results may thus indicate the presence
of a weak \textit{inverse} FIP effect in the corona of \object{AD
  Leo}. The value of carbon is too high to fit into the overall
picture of an inverse FIP effect. In our analysis the most recent
value of the \element{C} abundance in the solar photosphere
\citep{carbon} is used. By using the \element{C} abundance of
\citet{anders}, our value of \element{C}/\element{O} would shift
downwards by a factor of 1.5 and this abundance comes more in line
with the other high FIP elements.  

One possible effect for this high \element{C} abundance is that the EM
at low temperatures is not well constrained by the XMM-Newton
data. Therefore the \element{C} abundance, which is essentially formed
at low temperatures, is less certain. 

\section{Conclusions}
\label{sec:conclusion}

The physical properties of the corona of \object{AD Leo} are derived
from observations of two different satellites, XMM-Newton and
Chandra. The results from this analysis are in good agreement for both 
observations. 

One can distinguish for the flare state a broad range in temperatures
from about 1 to 40 MK, and for the quasi-steady state a range from 1
to 20 MK. The corresponding emission measures are not the same during
these states. The flare state has a higher emission measure for the
temperatures in the highest temperature range, which means that this
temperature has a higher contribution to the flare spectra than to the
spectra of the quasi-steady state.  

From the line triplets of some \element{He}-like ions the electron
temperatures and upper limits of the electron densities were derived,
which are representative for those plasmas in which \element{He}-like
ions are formed. These temperatures form only part of the range
presented by the DEM models. The results from this analysis show
higher electron temperatures during the flare state and also higher
upper limits of most of the electron densities, but these higher upper
limits may be due to different signal-to-noise ratios. 

The abundances, however, do not show a significant difference between
the flare state and the quasi-steady state. The derived average
abundances from the total XMM-Newton observation as well as for the
total Chandra observation are plotted against their first ionization
potential. These figures suggest the presence of an \textit{inverse}
FIP effect in the corona of \object{AD Leo}. This is in contrast with
the \object{Sun}, but in agreement with other active stars (\object{HR
  1099}: \citet{ifip1}; \object{AB Dor}: \citet{ifip2}).  

\begin{acknowledgements}
The SRON National Institute for Space Research is supported
financially by NWO. MG and MA acknowledge support from the Swiss
National Science Foundation (grant 2000-058827 and fellowship
81EZ-67388, respectively). 

Based on observations obtained with XMM-Newton, an ESA science mission
with instruments and contributions directly funded by ESA Member
States and the USA (NASA). 
\end{acknowledgements}

\bibliographystyle{aa}
\bibliography{h4285}

\begin{thebibliography}{21}
\expandafter\ifx\csname natexlab\endcsname\relax\def\natexlab#1{#1}\fi

\bibitem[{Allende~Prieto {et~al.}(2001)Allende~Prieto, Lambert, \&
  Asplund}]{oxygen}
Allende~Prieto, C., Lambert, D.~L., \& Asplund, M. 2001, \apj, 556, L63

\bibitem[{Allende~Prieto {et~al.}(2002)Allende~Prieto, Lambert, \&
  Asplund}]{carbon}
---. 2002, \apj, 573, L137

\bibitem[{Anders \& Grevesse(1989)}]{anders}
Anders, E. \& Grevesse, N. 1989, \\ \gca, 53, 197

\bibitem[{Audard {et~al.}(2000)Audard, G\"udel, Drake, \&
  Kashyap}]{latetypeflare}
Audard, M., G\"udel, M., Drake, J.~J., \& Kashyap, V.~L. 2000, \apj, 541, 396

\bibitem[{Brinkman {et~al.}(2001)Brinkman, Behar, G\"udel, Audard, den
  Boggende, Branduardi-Raymont, Cottam, Erd, den Herder, Jansen, Kaastra, Kahn,
  Mewe, ans J.~R.~Peterson, Rasmussen, Sakelliou, \& de~Vries}]{ifip1}
Brinkman, A.~C., Behar, E., G\"udel, M., {et~al.} 2001, \aap, 365, L324

\bibitem[{Brinkman {et~al.}(2000)Brinkman, Gunsing, Kaastra, van~der Meer,
  Mewe, Paerels, Raassen, van Rooijen, Br\"auninger, Burkert, Burwitz, Hartner,
  Predehl, Ness, Schmitt, Drake, Johnson, Juda, Kashyap, Murray, Pease,
  Ratzlaff, \& Wargelin}]{chandra}
Brinkman, A.~C., Gunsing, C. J.~T., Kaastra, J.~S., {et~al.} 2000, \apj, 530,
  L111

\bibitem[{Cully {et~al.}(1997)Cully, Fisher, Hawley, \& Simon}]{nh}
Cully, S.~L., Fisher, G.~H., Hawley, S.~L., \& Simon, T. 1997, \apj, 491, 910

\bibitem[{den Herder {et~al.}(2001)den Herder, Brinkman, Kahn,
  Branduardi-Raymont, Thomsen, Aarts, Audard, Bixler, den Boggende, Cottam,
  Decker, Dubbeldam, Erd, Goulooze, G\"udel, Guttridge, Hailey, Janabi,
  Kaastra, de~Korte, van Leeuwen, Mauche, McCalden, Mewe, Naber, Paerels,
  Peterson, Rasmussen, Rees, Sakelliou, Sako, Spodek, Stern, Tamura, Tandy,
  de~Vries, Welch, \& Zehnder}]{rgs}
den Herder, J.~W., Brinkman, A.~C., Kahn, S.~M., {et~al.} 2001, \aap, 365, L7

\bibitem[{Feldman(1992)}]{fip}
Feldman, U. 1992, \physscr, 46, 202

\bibitem[{Grevesse \& Sauval(1999)}]{iron}
Grevesse, N. \& Sauval, A.~J. 1999, \aap, 347, 348

\bibitem[{G\"udel {et~al.}(2001)G\"udel, Audard, Briggs, Haberl, Magee, Maggio,
  Mewe, Pallavicini, \& Pye}]{ifip2}
G\"udel, M., Audard, M., Briggs, K., {et~al.} 2001, \aap, 365, L336

\bibitem[{G\"udel {et~al.}(2003)G\"udel, Audard, Kashyap, Drake, \&
  Guinan}]{flareheating2}
G\"udel, M., Audard, M., Kashyap, V.~L., Drake, J.~J., \& Guinan, E.~F. 2003,
  \apj, 582, 423

\bibitem[{Hawley {et~al.}(1995)Hawley, Fisher, Simon, Cully, Deustua,
  Jablonski, Johns-Krull, Pettersen, Smith, Spiesman, \& Valenti}]{uvopt}
Hawley, S.~L., Fisher, G.~H., Simon, T., {et~al.} 1995, \apj, 453, 464

\bibitem[{Kaastra {et~al.}(1996{\natexlab{a}})Kaastra, Mewe, Liedahl, Singh,
  White, \& Drake}]{dem}
Kaastra, J.~S., Mewe, R., Liedahl, D.~A., {et~al.} 1996{\natexlab{a}}, \aap,
  314, 547

\bibitem[{Kaastra {et~al.}(1996{\natexlab{b}})Kaastra, Mewe, \&
  Nieuwenhuijzen}]{spex}
Kaastra, J.~S., Mewe, R., \& Nieuwenhuijzen, H. 1996{\natexlab{b}}, in UV and
  X-ray Spectroscopy of Astrophysical and La\-bo\-ratory Plasmas, K. Yamashita,
  T. Watanabe (eds.), Universal Academy Press, Inc., Tokyo, 411, (SPEX)

\bibitem[{Kahn {et~al.}(1979)Kahn, Linsky, Mason, Haisch, Bowyer, White, \&
  Pravdo}]{kahn}
Kahn, S.~M., Linsky, J.~L., Mason, K.~O., {et~al.} 1979, \apj, 234, L107

\bibitem[{Kashyap {et~al.}(2002)Kashyap, Drake, G\"udel, \&
  Audard}]{flareheating}
Kashyap, V.~L., Drake, J.~J., G\"udel, M., \& Audard, M. 2002, \apj, 580, 1118

\bibitem[{Mewe {et~al.}(1995)Mewe, Kaastra, \& Liedahl}]{mekal}
Mewe, R., Kaastra, J.~S., \& Liedahl, D.~A. 1995, Legacy 6, 16, (MEKAL)

\bibitem[{Pallavicini {et~al.}(1989)Pallavicini, Tagliaferri, \&
  Stella}]{exosat}
Pallavicini, R., Tagliaferri, G., \& Stella, L. 1989, \aap, 228, 403

\bibitem[{Pettersen {et~al.}(1984)Pettersen, Coleman, \& Evans}]{flare}
Pettersen, B.~R., Coleman, L.~A., \& Evans, D.~S. 1984, \apjs, 54, 375

\bibitem[{Porquet {et~al.}(2001)Porquet, Mewe, Dubau, Raassen, \&
  Kaastra}]{porquet}
Porquet, D., Mewe, R., Dubau, J., Raassen, A. J.~J., \& Kaastra, J.~S. 2001,
  \aap, 376, 1113

\end{thebibliography}

\end{document}